\documentclass[aps,preprintnumbers, supercriptaddress, amsmath, showpacs]{revtex4}
\usepackage{psfrag}
\usepackage{epsfig}
\usepackage{amsfonts} 
\usepackage{amssymb}  
\usepackage{graphicx} 
\usepackage{amsmath}  
\usepackage{amsmath}  

\usepackage[latin1]{inputenc}
\usepackage{color}
\usepackage{suffix}
\usepackage{mathtools}

\newcommand{\bea}{\begin{eqnarray}}
\newcommand{\eea}{\end{eqnarray}}
\newcommand{\beas}{\begin{eqnarray*}}
\newcommand{\eeas}{\end{eqnarray*}}

\def\Title#1{\begin{center} {\Large {\bf #1} } \end{center}}

\begin{document}

\Title{Quark matter contribution to the heat capacity of magnetized neutron stars}

\author{E. J. Ferrer, V. de la Incera and P.  Sanson}
\affiliation {Department of Physics and Astronomy, University of Texas Rio Grande Valley, 1201 West University Drive, Edinburg, Texas 78539, USA }


\begin{abstract}

In this paper, we find the heat capacity of the magnetic dual chiral density wave (MDCDW) phase of dense quark matter and use it to explore the feasibility of this phase for a neutron star interior. MDCDW is a spatially inhomogeneous phase of quark matter known to be favored at intermediate densities over the chirally symmetric phase and the color-flavor-locked superconducting phase. By comparing our result to the lower limit of the core heat capacity established from observations of transiently accreting neutron stars, we show that the heat capacity of MDCDW quark matter is well above that lower limit and hence cannot be ruled out. This result adds to a wealth of complementary investigations, all of which has served to strengthen the viability of a neutron star interior made of MDCDW quark matter. For completeness, we review the contributions to the heat capacity of the main neutron star ingredients at low, high and intermediate densities, with and without the presence of a magnetic field. 
\end{abstract}


\maketitle

\section{Introduction}

The composition of the inner phase of neutron stars (NS) remains as a challenging open
question in astrophysics. Depending on the density reached at the core, different phases could be
realized. Close to the nuclear saturation density, $\rho_s=2.8 \times 10^{14}$ g/cm$^3$, nuclear-star matter results in almost pure neutron matter with a very small
proton fraction, which is electrically neutralized with a corresponding density of electrons (see \cite{NS-Rev}
for review). The degenerate neutron system will form a superfluid due to the attractive interaction mediated by pion exchange between two neutrons on the Fermi surface with zero total momentum
\cite{Superfluidity}. On the other hand, this nucleon-nucleon attractive interaction will also be responsible for the
formation of protons' Cooper pairs creating an electric superconductor. The expected density in the
cores of massive NSs could be even larger [$\sim (5 - 10) \rho_s$]. In those highly dense cores, the
neutron-rich matter can give rise to more exotic degrees of freedom, like hyperons \cite{Hyperons}, and perhaps
even transitioning to a quark matter phase (see \cite{Quark-Matter} for review) forming what is known as a hybrid star.
Besides, since cold strange quark matter could be absolutely stable \cite{Witten}, a phase transition may occur
that would favor the creation of a quark-matter phase over the entire star, thus giving rise to a strange star.

The ground state of the superdense quark-matter system is unstable with respect to the formation
of diquark condensates \cite{CS}, a nonperturbative phenomenon essentially equivalent to the Cooper
instability just mentioned for protons. Given that in QCD one-gluon exchange between two quarks
is attractive in the color-antitriplet channel, quarks should condense into Cooper pairs, which are
color antitriplets. These color condensates break the SU(3) color gauge symmetry of the ground
state producing a color superconductor. At densities much higher than the masses of the u, d, and s
quarks, one can assume the three quarks as massless. In this asymptotic region the most favored state
is the so-called color-flavor-locked (CFL) phase \cite{CFL}, characterized by a spin-zero diquark condensate
antisymmetric in both color and flavor.

Nonetheless, the densities at NS cores are not asymptotically large. At more realistic, intermediate densities, other phases different from the CFL can be realized. It has long been argued that in this region of intermediate density and relatively low temperature, particle-hole condensates, in which the pairs carry a net total momentum, are preferable, giving rise to spatially inhomogeneous chiral condensates. Spatially inhomogeneous chiral phases have been theoretically found in the large-N limit of QCD \cite{Large-N}, 
in Nambu-Jona-Lasinio (NJL) models \cite{NJL}-\cite{MDCDW-2}, and in quarkyonic matter \cite{Quarkyonic}.

Another element that cannot be ignored when studying the physics of NS is their magnetic fields. NS typically have surface magnetic fields of the order of $10^{12}$ G, or even larger ($10^{14}-10^{15}$ G) in the case of magnetars \cite{Magnetar}. Core field magnitudes are stronger, since the magnetic flux is conserved thanks to the high electric conductivity of stellar matter, settling a larger magnetic field in the denser central region of the star. Inner field estimates based on the viral theorem range from $10^{18}$ G for nuclear matter \cite{Nuclear-matter-field} to $10^{20}$ G for quark matter \cite{Fermions-B, Laura}.  Nevertheless, when the hydrodynamic equilibrium between gravity and matter pressures has been taken into account following different approaches \cite{Cardall, Aric},  the resultant maximum field value for stable configurations has been of order $eB \sim 10^{17}$ G.

Different studies have shown that large magnetic fields hardly affect the equation of state (EOS) of a NS \cite{Aric, Laura, Aurora, Estefano} except when it has very significant magnitude. Nevertheless, depending on the phase, moderate magnetic fields can produce interesting qualitative effects. For example, in color superconductivity, the CFL phase exhibits three different phases with increasing magnetic field \cite{vortices}. At intermediate densities, moderate magnetic fields enable the appearance of a topological fermion structure in a density wave chiral phase \cite{Klimenko} that ultimately leads to the anomalous transport properties of the so-called magnetic dual chiral density wave (MDCDW) phase \cite{MDCDW}.

So far, connecting the EOS of a given matter phase to the star mass and radius has failed in part because, from the observational side, the precision in the radius determination is not yet accurate enough, and because, from the theoretical side, several uncontrolled parameters of the effective models considered can play a decisive role in the behavior of the EOS, while more trustworthy nonperturbative methods, as lattice calculations, are not applicable at finite densities. 

Another direction, that can help to discriminate among potential phases, aims to connect the transport properties of different matter phases with the observed behavior of NS cooling. Soon after a supernova collapses a young NS emerges with $T \sim 10^{11}$ K, which cools mainly through two steps: First, it cools by neutrino emission, and then it cools down to $T \sim 10^8$ K by photon emission from the surface \cite{Temp-NS}. In a first approximation, this scenario can be grasped from a simple energy-balance equation
\begin{equation} \label{Ener-Bal-Eq}
\frac{dT}{dt}=-\frac{L_\nu +L_\gamma-H}{C_v},
\end{equation}
where $T$ is the inner medium temperature, $t$ the star's age, $C_v$ is the specific heat at constant volume, $L_\nu$ and $L_\gamma$ are neutrino and surface photon luminosities, respectively; and $H$ is the heating rate associated with any additional heating source. Depending on the star's age, different components of the right-hand-side of Eq. (\ref{Ener-Bal-Eq}) will play major roles. But as seen from  (\ref{Ener-Bal-Eq}), the specific heat of the matter phase prevailing in the star core will always play a fundamental  role in the star thermal evolution. Being inversely proportional to the variation of temperature in time, having a medium with a high $C_v$ implies that the cooling process will be slow. 

It is worthy of notice that although old NS are relatively cold (i.e. with $T \sim 10^8 K \ll T_{Fermi} \sim 10^{12} K$), having temperatures that are negligibly smaller than the core  density, their
thermal properties can have a say in constructing a model for their interiors \cite{Cv-NS, Cv-NS-2}. For example, assuming that NS cores cool completely between outbursts, it was found in \cite{Cv-NS} that according to observations 
of the temperatures of accreting NS in quiescence, the heat capacity of the NS core presents a lower limit given by $\tilde{C}_V\gtrsim 10^{36}(T/10^8)$ erg/K. Hence, matter phases that do not satisfy this lower-limit constraint should be ruled out. For example, any superfluid/superconducting phase will be discarded. In particular, as was argued in \cite{Cv-NS}, if the quark phase that will prevail at the core densities is the CFL phase, it will imply that quarks cannot be a suitable choice for the inner composition of compact stars. First, because being the CFL a color superconductor, its quark's heat capacity will be strongly depleted, as we will discuss in more detail in this paper, and second, because its neutrality is guaranteed by its $u$, $d$ and $s$ quark components alone,  without the necessity to have electrons. Thus, instead of a quark inner phase, a nuclear phase with a larger electron fraction will be preferred.

 In this paper, we will show that a quark phase  (i.e. the MDCDW phase mentioned above) that is more appropriate at the intermediate densities that are likely to prevail in NS, can satisfy the lower-limit constraint with a value $\tilde{C}_V\simeq 10^{38}(T/10^8)$ erg/K, redeeming the possibility to have quark matter as a constituent of the inner core. As we will show, the quark contribution to the heat capacity of this phase is linear in $T$ and in $eB$, as it is for electrons in a magnetic field, but with a heat capacity that is 1 order larger than that of electrons under the same conditions. Moreover, if in this intermediate-density quark phase only the $u$ and $d$ flavors participate, the neutrality condition requires the presence of electrons in addition to the quarks \cite{Estefano}. Thus, the electrons will also contribute to the system heat capacity. 

The paper is organized into three main sections where the contributions to the heat capacity of the main NS ingredients at low, high and intermediate densities are studied making special emphasis on the effect produced by a magnetic field.
Thus, in Section II, for the sake of unity and consistency, we recapitulate the calculations of the heat capacity of relativistic electrons at low temperature, as well as for superfluid neutrons. We also include the effect on the electron's $C_V$ of an applied magnetic field. In Section III, we calculate the heat capacity at low temperature of  quark matter first in the CFL phase, and then in the MCFL phase, where an applied magnetic field plays an important role \cite{vortices}. The $C_V$ of NS at intermediate densities is investigated in Section IV through the MDCDW phase. In this phase, the presence of a magnetic field is essential to guarantee the stability with respect to thermal fluctuations \cite{MDCDW-2}.  In Section V, the leading contributions to the heat capacity of NS of the main components of the different inner phases are estimated.
Finally, in Section VI, we make our concluding remarks deepening in the relation of our results at intermediate densities and the astrophysics of NS. In the Appendix, we study the neutrality condition for an electron-proton system, to analyze the relation between the electric and baryonic chemical potentials with the electron mass, which is a result used in other sections.

\section{$C_V$ of neutron stars at low density}

For baryon densities not too much larger than saturation density, it is expected that nuclear matter will prevail in the NS inner core. Then, the main components we will consider will be neutrons, protons and a certain amount of electrons that will ensure the electric neutrality of the stellar medium. Thus, in this section we review the calculation of the heat capacity of these components in the low-temperature limit  (i. e. for $T \ll \mu_e, \mu, m_e$, with $\mu_e$ and $\mu$ denoting the electric and  baryonic chemical potentials respectively and $m_e$ denoting the electron mass). Hence, in determining the thermal behavior of these media, we should be aware that because of its low temperature and high density, quantum-relativistic statistics plays a crucial role in the thermodynamic and transport properties of NS.

As known, the heat capacity can be calculated from the thermodynamic potential, $\Omega$, of the thermal bath of the particles under study. In general, it is given by 
\begin{equation}
    C_V= -T\frac{\partial^2\Omega}{\partial T^2}=-2\beta^2\frac{\partial\Omega}{\partial \beta}+\beta^3 \frac{\partial^2 \Omega}{\partial \beta^2},
    \label{sh}
\end{equation}
where $\beta$ is the inverse of the absolute temperature.

\subsection{$C_V$ of relativistic electrons at low temperature}

In the case of electrons, the NS inner magnetic field is large enough to have a noticeable effect.  Therefore, we will calculate the electron's $C_V$  at zero magnetic field first, and then in the presence of a magnetic field to compare the two results.

\subsubsection{Electrons at $B = 0$}

In order to calculate $C_V$ using formula (\ref{sh}), only the statistical part of the thermodynamic potential matters. After performing the Matsubara's frequency sum, the electron thermodynamic potential in the one-loop approximation at finite temperature and density is given by the well-known result \cite{Dolan}
\begin{equation}
    \Omega_\beta^{e} = -\frac{2}{\beta}\int_{-\infty}^{\infty}\frac{d^3p}{(2\pi)^3}\left[ln(1+ e^{-\beta(\epsilon_e+\mu_e)})+ln(1+e^{-\beta(\epsilon_e-\mu_e)})\right],
    \label{tp_e}
\end{equation}
where the energy spectrum is $ \epsilon_e=\sqrt{p^2+m_e^2}$. The numerical factor, 2, comes from the spin degeneracy.

Substituting Eq. (\ref{tp_e}) into (\ref{sh}), we obtain
\begin{equation}
    C^{e}_V = \frac{1}{16\pi^3T^2}\int_{-\infty}^{\infty}d^3p\left[(\epsilon_e+\mu_e)^2 sech^2\left(\frac{\epsilon_e+\mu_e}{2T}\right)+(\epsilon_e-\mu_e)^2 sech^2\left(\frac{\epsilon_e-\mu_e}{2T}\right)\right]
    \label{CVelectrons}
\end{equation}

Assuming that $\mu_e > 0$, we can discard, in the low-temperature limit,  the positron's contribution in (\ref{CVelectrons})
\begin{equation}
    C^{e}_V \simeq \frac{1}{4\pi^2T^2}\int_{0}^{\infty}dp p^2(\epsilon_e-\mu_e)^2sech^2\left(\frac{\epsilon_e-\mu_e}{2T}\right)
\end{equation}
Now, using the dispersion relation, we can make a variable change to integrate in energy 
\begin{equation}
    C^{e}_V \simeq \int_{m_e}^{\infty}d\epsilon_e g(\epsilon)\left(\frac{\epsilon_e-\mu_e}{2T}\right)^2sech^2\left(\frac{\epsilon_e-\mu_e}{2T}\right)
    \label{Cv-T-B}
\end{equation}
where we introduce the density of state per unit volume function 
\begin{equation}
    g(\epsilon_e)=\frac{\epsilon_e\sqrt{\epsilon_e^2-m_e^2}}{\pi^2}
    \label{State-Func}
\end{equation}

As it is known, the Fermi-Dirac distribution for fermions at low temperatures (i.e. when $k_B T < \epsilon_F$, with $k_B$ denoting the Boltzmann constant and $\epsilon_F=\mu_e$, the Fermi energy), 
only changes significantly in the vicinity of $\epsilon_F$. Hence, the integral in (\ref{Cv-T-B}) mainly gets
contributions in the vicinity of $\epsilon_F$ and we can approximate the density
of states $g(\epsilon)$ with its value at $g(\epsilon_F )$. Thus, after this change and introducing the new variable $x=(\epsilon_e-\epsilon_F)/T$, we write (\ref{Cv-T-B}) as
\begin{equation}
    C^{e}_V\simeq  g(\epsilon_F)T\int_{\frac{m-\epsilon_F}{T}}^{\infty}\left (\frac{x}{2}\right )^2sech^2(\frac{x}{2})dx
\end{equation}
For $m_e<\mu_e$ (check in the Appendix that this inequality is in agreement with the electric neutrality condition), we have that $\lim_{T \to 0}(\frac{m_e-\epsilon_F}{T}) \to -\infty$, then
\begin{eqnarray}
        C^{e}_V&\simeq& 2g(\epsilon_F)T\int_{0}^{\infty}\left (\frac{x}{2}\right )^2sech^2(\frac{x}{2})dx
    \\  
        &\simeq& \frac{\pi^2}{3}g(\epsilon_F)T
\end{eqnarray}

Finally, we obtain
\begin{equation}
    C^e_V\simeq \frac{ \mu_eT}{3}\sqrt{\mu_e^2-m_e^2}
    \label{Cv-electron}
\end{equation}

\subsubsection{Electrons at $B \neq 0$}

The statistical part of the electron thermodynamic potential in the presence of a constant and uniform magnetic field is given by \cite{Fermions-B}
\begin{equation}
    \Omega_\beta^{e} = -\frac{eB}{4\pi^2\beta}\int_{-\infty}^{\infty}dp\sum_{n=0}^\infty d(n)\left[ln(1+ e^{-\beta(\epsilon_n+\mu_e)})+ln(1+e^{-\beta(\epsilon_n-\mu_e)})\right],
    \label{Omega-B}
\end{equation}
where the field-dependent energy spectrum is $\epsilon_n^2=p^2+2|eB|n+m_e^2$, the degeneracy is given by $d(n)=2-\delta_{n0}$ with $n$ denoting the Landau level numbers n=0,1,2,....

In the low-temperature limit, the leading contribution of (\ref{Omega-B}) reduces to
\begin{equation}
    \Omega_{(\beta-LLL)}^{e} \simeq -\frac{eB}{4\pi^2\beta}\int_{-\infty}^{\infty} ln(1+e^{-\beta(\epsilon_0-\mu_e)})dp
    \label{Omega-B-2}
\end{equation}
with only the particles confined into the lowest Landau level (LLL)  participating.

Making a variable change from momentum to energy, we have
\begin{equation}
    \Omega_{(\beta-LLL)}^{e} \simeq -\frac{2}{\beta}\int_{m_e}^{\infty}d\epsilon_0 g_B(\epsilon_0)ln[1+e^{-\beta(\epsilon_0-\mu_e)}]
\label{Cv}
\end{equation}
Here, we  introduce the density of state per unit volume of the LLL, 
\begin{equation}
   g_B(\epsilon_0)=\frac{eB\epsilon_0}{4\pi^2\sqrt{\epsilon_0^2-m_e^2}}
\end{equation}
Using formula (\ref{sh}), with the thermodynamic potential (\ref{Cv}), we obtain the corresponding heat capacity
\begin{equation}
 C_V^e(B)\simeq 2 \int_{m_e}^{\infty}d\epsilon_0 g(\epsilon_0)\left(\frac{\epsilon_0-\mu_e}{2T}\right)^2 sech^2\left(\frac{\epsilon_0-\mu_e}{2T}\right)
\end{equation}
Following the same steps as in the last section at $B=0$, we obtain in the low-temperature limit 
\begin{equation}
     C_V^e(B) \simeq 2g(e_F)T\int_{-\infty}^{\infty}\left (\frac{x}{2} \right )^2sech^2(\frac{x}{2})dx =  \frac{2\pi^2}{3}g(\epsilon_F)T
\end{equation}

Therefore
\begin{equation}
     C_V^e(B)\simeq \frac{eB\mu_eT}{6\sqrt{\mu_e^2-m_e^2}}
\label{Cv-LLL}
\end{equation}


Here, the following comment is in order. In considering the effect of magnetic fields in the electron heat capacity, we neglected the contribution of the interaction of the field with the anomalous magnetic moment of the electron. It is based on the results of Ref. \cite{Aurora}, where it was shown that this contribution in a strong, field as well as in weak fields, produces a negligible effect on the thermodynamics of charged fermions.

Finally, considering that $\mu_e > m_e$ (see the Appendix), we have from (\ref{Cv-LLL}), that the leading contribution of the electron heat capacity in the presence of a magnetic field is given by $C_V^e(B)\simeq eBT/6$.

\subsection{$C_V$ of a nonrelativistic  superfluid of neutrons}

Let us suppose that the baryonic chemical potential is not so much higher than the neutron mass. In this case, the nonrelativistic approximation is the more suitable one. As we discussed in the Introduction, thanks to the attractive pion exchange, the neutrons close to the Fermi surface can form pairs that at low temperatures can produce a boson condensate. Since neutrons are electrically neutral, the  pair condensates will keep the neutrality of the ground state; thus, they will form a  macroscopic superfluid state (In the case of protons, since they are electrically charged, their Cooper pairs are charged, and their condensation creates a superconductor).

For a non-relativistic superfluid the thermodynamic potential arises from the thermally excited quasiparticles, which following the Landau approach \cite{Landau}, are treated as an ideal Bose gas, 
\begin{equation}
    \Omega^{n}_\beta = \frac{1}{\beta}\int\frac{d^3p}{(2\pi)^3}\ln \left(1-e^{-\beta\epsilon_n}\right )
    \label{Therm-N}
\end{equation}
with a gapped energy spectrum,
\begin{equation}
   \epsilon_n=\frac{(p-p_0)^2}{2m}+\Delta
    \label{Energy-N}
\end{equation}
Here, $\Delta$ is the energy gap created by the neutron-pair condensates, and $p_0$ is the gap location in momentum space.

 Substituting (\ref{Therm-N}) into (\ref{sh}), we find the heat capacity
\begin{equation}
    C^{n}_V = \frac{1}{2\pi^2T^2}\int_0^{\infty}dp p^2 e^{(-\epsilon_n/T)}\epsilon_n^2
\end{equation}

Introducing the explicit form of the dispersion relation (\ref{Energy-N}), we have
\begin{equation}
    C^{n}_V = \frac{ e^{-(\Delta/T)}} {2\pi^2T^2}\int_0^{\infty}dp p^2 e^{-(p-p_0)^2/2mT}\left(\Delta+\frac{(p-p_0)^2}{2m}\right)^2
\end{equation}

Making the  variable change $x=(p-p_0)/\sqrt{2mT}$, we obtain
\begin{equation}
    C^{n}_V =\frac{ e^{-(\Delta/T)}\sqrt{2mT}}{2\pi^2T^2}\int_{-\frac{p_0}{\sqrt{2mT}}}^\infty dx \left [p_0^2+2x\sqrt{2mT}p_0+2mTx^2\right]e^{-x^2}(\Delta+x^2T)^2
\end{equation}
In the low-temperature limit, the previous equation can be simplified as
\begin{equation}
    C^{n}_V \simeq \frac{e^{-(\Delta/T)}\sqrt{2m}p_0^2\Delta^2}{2\pi^2T^{\frac{3}{2}}}\int_{-\infty}^{\infty}dx e^{-x^2}
\end{equation}
where higher-order terms in $T/\Delta$ are neglected.

Finally, we obtain
\begin{equation}
    C^{n}_V \simeq \frac{e^{-(\Delta/T)}\Delta^2}{T^{\frac{3}{2}}}\sqrt{\frac{m}{2\pi^3}}p_0^2
    \label{Cv-N-Final}
\end{equation}

From (\ref{Cv-N-Final}), we can see the known result that the $C_V$ of superfluids is exponentially damped by the gap. 

The effect of magnetic fields on the specific heat of a free gas of neutrons was investigated in \cite{Aric}. In this case the magnetic field can interact only through the anomalous magnetic moment of the neutrons. There, it was found that up to magnetic fields of $10^{18}$ G, the magnetic field effect on $C_V$ is negligibly small. In the case of superfluidity, since the gap is neutral and formed by neutral particles (see in Section III, that in color superconductivity although the gap is neutral with respect to the rotated electromagnetism, its inner composition given by charged quarks makes a difference), the leading term in (\ref{Cv-N-Final}), $e^{-(\Delta/T)}$, remains  unaffected.

\subsection{$C_V$ of superfluid phonons}
The existence of a superfluid of neutrons at $T\leqslant T_c=10^{10}$ K gives rise to another participant in the heat transfer of a NS- the superfluid phonon. The phonon in this context is generated as a massless Goldstone mode by the breaking of the baryonic symmetry produced by the condensation of the s-wave neutron Cooper pairs. This phenomenon was investigated in Ref. \cite{Superfluid-Phonon} and the corresponding specific heat was found to be
\begin{equation}\label{SuperfluidPhonon}
    C^{(sPh)}_V =\frac{2\pi^2T^3}{15v_s^3},
\end{equation}
where $v_s\simeq k_{Fn}/\sqrt{3}M$, with $M$ being the neutron mass and $k_{Fn}$ the neutron Fermi momentum.

At low temperatures, the electron contribution to $C_V$, (\ref{Cv-electron}), is larger than that of the superfluid phonons (\ref{SuperfluidPhonon}). hence, the electron contribution to $C_V$ is the dominant one. Nevertheless, it is worthy of mention that in the presence of a 
moderate to high magnetic field ($B \geqslant 10^{13}$ G) it was found in \cite{Superfluid-Phonon} that the contribution to the heat conduction of the superfluid phonons is larger than that of the electrons in the direction transverse to the field. This is due to the fact that in the presence of a magnetic field the electron heat transport is very anisotropic with the electron motion in the transverse direction limited by the Landau quantization, while the phonons, being neutral, can equally move along and transverse to the field directions.

It is important to notice that at high densities, the s-wave interactions between neutrons become repulsives, while the attractive p-wave ones, which produce spin-1 Cooper pairs, are favored  \cite{p-wave-condensate-1, p-wave-condensate-2}. This condensate breaks rotational and baryonic symmetries and gives rise to two gapless Goldstones (called {\it angulons}) associated with the breaking of the rotational symmetry with respect to the two perpendicular axes to the condensate spin direction and to an extra Goldstone, the superfluid phonon. In this case, each of the Goldstone modes will contribute to $C_V$ with a term similar to Eq.  (\ref{SuperfluidPhonon}) \cite{p-wave-condensate-2}.

\section{$C_V$ of neutron stars at high density}

Let us consider that the core density is high enough to deconfine quarks. Under such conditions, since the quarks sitting on the Fermi sphere will suffer the Cooper instability produced by the attractive one-gluon exchange interaction, they will condense in color superconducting pairs. If the density is even larger than the mass of the $s$ quark, the quark Cooper pairs will form the most energetically favored CFL phase \cite{CFL}. 

As follows, we consider the heat capacity of each of the main contributions present in the CFL phase  \cite{CFL-Omega}, at zero magnetic field first, and then, in the presence of a magnetic field, which forms the so-called magnetic CFL (MCFL)  phase \cite{MCFL}. 

\subsection{$C_V$ of quark matter in the CFL phase}

In the CFL phase, the temperature-dependent thermodynamic potential is \cite{CFL-Omega}
\begin{equation}
    \Omega_\beta^{CFL} = -\frac{1}{2\pi^2\beta}\sum_{i=1}^4\int_{0}^{\Lambda}dpp^2ln[1+e^{-\beta|\epsilon_i|}],
    \label{Ther-CFL}
\end{equation}
with energy spectra given by
\begin{equation}
   |\epsilon_{1,2}|=\sqrt{(p\mp\mu)^2+\Delta^2}, \qquad |\epsilon_{3,4} |= \sqrt{(p\mp\mu)^2+4\Delta^2}
\end{equation}
where $\Delta$ is the CFL gap, $\mu$ is the baryonic chemical potential and $\Lambda$ is the momentum cutoff of the low-energy theory described by a NJL model \cite{CFL}.

Corresponding to the thermodynamic potential (\ref{Ther-CFL}) we have the heat capacity
\begin{equation}
    C^{CFL}_V = \frac{1}{2\pi^2T^2}\sum_{i=1}^4 \int_0^{\Lambda} dp p^2 \epsilon_i^2 sech^2\left(\frac{|\epsilon_i|}{2T}\right)  
\end{equation}

Doing the variable change $x=p/T$ we obtain
\begin{equation}
    C^{CFL}_V = \frac{T^3}{2\pi^2}\sum_{i=1}^4 \int_0^{\widehat{\Lambda}} dx x^2 \widehat{\epsilon}_i^2 sech^2\left(\frac{|\widehat{\epsilon}_i|}{2}\right) 
    \label{C-CFL} 
\end{equation}
where we use the notation $\widehat{Q}=Q/T$ for all the quantities, including $ |\widehat{\epsilon}_{1,2}|=\sqrt{(x\mp\widehat{\mu})^2+\widehat{\Delta}^2}$ and $|\widehat{\epsilon}_{3,4} |= \sqrt{(x\mp\widehat{\mu})^2+4\widehat{\Delta}^2}$.

In the low-temperature limit, the leading contribution in (\ref{C-CFL}) can be expressed as
\begin{equation}
    C^{CFL}_V \simeq \frac{T^3}{2\pi^2}\sum_{i=1}^4 \int_0^{\widehat{\Lambda}} dx x^2 \widehat{\epsilon}_i^2 \exp(-|\widehat{\epsilon}_i|)
    \label{C-CFL-2} 
\end{equation}

 Because of the negative exponential, the main contribution in (\ref{C-CFL-2})  comes from the region around the integral lower limit. Thus, up to a numerical coefficient, we can extract the leading contribution making
\begin{equation}
    C^{CFL}_V \simeq \left[ \frac{(\mu^2+\Delta^2)T}{\pi^2}e^{ -\sqrt{\mu^2+\Delta^2}/T}+\frac{(\mu^2+4\Delta^2)T}{\pi^2}e^{ -\sqrt{\mu^2+\Delta^2}/T} \right] \int_0^1 dx x^2
    \label{C-CFL-3} 
\end{equation}

Finally, we have
\begin{equation}
    C^{CFL}_V \simeq  \frac{(\mu^2+\Delta^2)T}{3\pi^2}e^{ -\sqrt{\mu^2+\Delta^2}/T}+\frac{(\mu^2+4\Delta^2)T}{3\pi^2}e^{ -\sqrt{\mu^2+\Delta^2}/T}
    \label{C-CFL-3} 
\end{equation}
Hence,  we obtain that in this case, similarly to the superfluid state, the heat capacity is exponentially damped at low temperature, with a damping factor that depends on the gap and baryonic chemical potential.

\subsection{$C_V$ of Goldstone modes in the CFL phase}

Taking into account that the symmetry-breaking pattern in the CFL is given
by \cite{Goldstone-CFL},
\begin{equation}\label{CFL}
\mathcal{G}=SU(3)_C \times SU(3)_L \times SU(3)_R\times
U(1)^{(1)}_A \times U(1)_B
\rightarrow SU(3)_{C+L+R}.
\end{equation} 
we have that this symmetry reduction leaves ten Goldstone bosons: a singlet
associated with the breaking of the baryonic symmetry $U(1)_B$, another singlet associated to the breaking of the $U(1)_A^{(1)}$ approximated symmetry [the group $U(1)^{(1)}_A$, not to be confused with the usual
anomaly $U(1)_{A}$, is related to the current which is an
anomaly-free linear combination of $s$, $d$, and $u$ axial
currents \cite{miransky-shovkovy-02}, and
an octet associated with the axial $SU(3)_A$ group.

Actually, if the baryonic chemical potential is not sufficiently large to justify neglecting the quark masses, the breaking of the axial  $SU(3)_A$ group is only apparent. In this case, we end up with an octet of massive pseudo-Goldstone modes associated with the breaking of the global symmetry $SU(3)_A$, another pseudo-Goldstone associated with the breaking of the $U(1)_A^{(1)}$ group, and only one massless Goldstone mode associated with the breaking of the baryonic symmetry. The corresponding $C_V$  of the massive and massless modes were found in Ref. \cite{Igor} and are given, respectively by
\begin{equation}
    C_V^B\simeq \frac{m^{7/2}}{2 \sqrt{2}\pi^{3/2}v^3 \sqrt{T}}e^{-m/T}   
  \label{Massive-G}
\end{equation}
and
\begin{equation}
    C_V^G = \frac{2\pi^2T^3}{15 v^3}
  \label{Massless-G}
\end{equation}
In Eqs. (\ref{Massive-G}) and (\ref{Massless-G}), $m$ is the mass of the pseudo-Goldstones and $v$ their velocity, which should be found from the CFL microscopic theory. Comparing (\ref{Massless-G}) with (\ref{C-CFL-3})  and (\ref{Massive-G}), we see that the contribution of the massless Goldstones is the dominant one in the CFL phase.

\subsection{$C_V$ of quark matter in the MCFL phase}

In the presence of a magnetic field the symmetries of the CFL phase are reduced, first, because a magnetic field interacting with quarks of different  electric charges reduces the flavor symmetry, and second, because a magnetic field breaks rotational symmetry.
The new phase was initially studied in Ref. \cite{MCFL} and was called the magnetic-color-flavor-locked phase (MCFL).

Here, we should mention that electromagnetism inside these color superconducting phases has to be redefined. 
Even though the original electromagnetic $U(1)_{em}$ symmetry is broken by the formation of quark Cooper pairs that are electrically charged \cite{Bailin}, a residual $\widetilde{U}(1)$ symmetry still remains \cite{CFL-Rotated-1}. The massless gauge field associated with this symmetry is given by the linear combination of the conventional photon field and the eighth-gluon field \cite{CFL-Rotated-1, CFL-Rotated-2}, $\widetilde{A}_{\mu}= \cos \theta A_{\mu}-\sin \theta G^8_{\mu}$ with the mixing angle given as a function of the strong coupling constant $g$ and the electromagnetic coupling $e$ as $\theta=\cos^{-1}(g/\sqrt{e^2/3+g^2})$. The field $\widetilde{A}_\mu$ plays the role of an in-medium or rotated electromagnetic field. Therefore, a magnetic field associated with $\widetilde{A}_\mu$ can penetrate the color superconductor without being subject to the Meissner effect, since the color condensate is neutral with respect to the corresponding rotated electric charge.

The temperature-dependent part of the thermodynamic potential of this phase is expressed through the contributions of the rotated-charged ($\Omega_C$) and rotated-neutral ($\Omega_N$) quarks \cite{Laura}
\begin{equation}
    \Omega_\beta^{MCFL} = \Omega_C+\Omega_N,
    \label{Ther-CFL}
\end{equation}
where
\begin{equation}
    \Omega_C = -\frac{\widetilde{e}\widetilde{B}}{4\pi^2\beta} \sum_{n=o}^\infty d(n)\sum_{n=1}^2 \int_0^\Lambda dp \ln \left (1+e^{-\beta |\epsilon_i^{(c)}|} \right )
    \label{Ther-CFL-C}
\end{equation}

\begin{equation}
    \Omega_N = -\frac{1}{4\pi^2\beta} \sum_{i=1}^6 \int_0^\Lambda dp p^2 \ln \left (1+e^{-\beta |\epsilon_i|} \right )
    \label{Ther-CFL-C}
\end{equation}
In (\ref{Ther-CFL-C}), $n=0,1,2,...$ denotes the Landau level number, $\widetilde{B}$ is the rotated magnetic field and the energy spectra are given by
\begin{equation}
    |\epsilon^{(c)}_{1,2}|=\left[ \left (\sqrt{p^2+2\tilde{e} \tilde{B}n}\pm\mu\right)^2+\Delta_H^2 \right ]^{1/2}
    \label{E-C}
\end{equation}

\begin{equation}
    |\epsilon^{(0)}_{1,2}|=\sqrt{(p\pm \mu)^2+\Delta^2 }, \quad |\epsilon^{(0)}_{3,4}|=\sqrt{(p\pm \mu)^2+\Delta_a^2} , \quad |\epsilon^{(0)}_{5,6}|=\sqrt{(p\pm \mu)^2+\Delta_b^2}
    \label{E-N}
\end{equation}

with
\begin{equation}
   \Delta_a=\frac{1}{2}\left[ \Delta+\sqrt{\Delta^2+8\Delta_H^2} \right] \quad \Delta_b=\frac{1}{2}\left[ \Delta-\sqrt{\Delta^2+8\Delta_H^2} \right] 
    \label{Gaps}
\end{equation}

In these expressions $\Delta$ and $\Delta_H$ denote the two gaps of the MCFL phase \cite{MCFL}. As seen, this phase, having a lower symmetry, doubles the number of gaps as compared with the CFL phase. Here, it is timely to comment that in the presence of a magnetic field, there is a third gap associated with the interaction of the magnetic field with the anomalous magnetic moment of the diquark pairs \cite{Bo}. This is a consequence of the breaking of the rotational symmetry by the uniform magnetic field, which opens new interaction channels. This new gap becomes significant only at very high magnetic fields  (i.e. $\sim 10^{19}$ G)  \cite{Bo}. Thus, in a moderate field, it can be neglected, and this is why we are not considering its effect here. 

The calculation of the heat capacity corresponding to $\Omega_N$ is equivalent to that of the CFL phase. Thus, as in (\ref{C-CFL-3}), its result will be exponentially damped at low temperature. Now, for the heat capacity corresponding to $\Omega_C$, we should notice that in the low-temperature limit, because of the negative exponential in (\ref{Ther-CFL-C}), the leading contribution to $C_V$, in amoderate magnetic field, comes from the LLL, as in the magnetized electron system. Hence, we will find the corresponding leading contribution to the heat capacity starting from
\begin{equation}
    \Omega^{(LLL)}_C \simeq -\frac{\widetilde{e}\widetilde{B}}{4\pi^2\beta} \sum_{n=1}^2 \int_0^\Lambda dp \ln \left (1+e^{-\beta |\epsilon_i^{(LLL)}|} \right )
    \label{Ther-CFL-C-2}
\end{equation}
with
\begin{equation}
    |\epsilon^{(LLL)}_{1,2}|=\sqrt{(p\pm\mu)^2+\Delta_H^2}
    \label{E-C-LLL}
\end{equation} 

From Eq. (\ref{Ther-CFL-C-2}), we obtain the heat capacity
\begin{equation}
      C^{MCFL}_V =- \frac{\widetilde{e}\widetilde{B}}{16\pi^2 T^2} \sum_{n=1}^2 \int_0^\Lambda dp |\widehat{\epsilon}_i|^2 sech^2\left( \frac{|\widehat{\epsilon}_i|}{2T}\right),
      \label{MCFL-C-Cv}
\end{equation}
where we are using the notation $\hat{Q} = Q/T$ for all physical quantities.

In the low-temperature limit, we can simplify (\ref{MCFL-C-Cv}) as
\begin{equation}
      C^{MCFL}_V \simeq\frac{\widetilde{e}\widetilde{B}}{4\pi^2 T^2} \sum_{n=1}^2 \int_0^\Lambda dp |\epsilon_i|^2e^{-|\epsilon_i|/T}
\end{equation}
Making the variable change $x=p/T$, we obtain 
\begin{equation}
    C^{MCFL}_V \simeq\frac{\widetilde{e}\widetilde{B}T}{2\pi^2} \int_0^1 \left(\hat{\mu}^2+\hat{\Delta}_H^2\right)e^{-\sqrt{ \hat{\mu}^2+\hat{\Delta}_H^2}} dx
\end{equation}
and finally,
\begin{equation}
    C^{MCFL}_V \simeq\frac{\widetilde{e}\widetilde{B}}{2\pi^2T} \left(\mu^2+\Delta_H^2\right)e^{-\sqrt{ \mu^2+\Delta_H^2}/T}
    \label{Cv-MCFL-T} 
\end{equation}

Comparing Eq. (\ref{Cv-MCFL-T}) with Eq. (\ref{C-CFL-3}), we see that the MCFL heat capacity is larger at low temperatures than that of the CFL phase. Despite this, the heat capacity in the MCFL case is also exponentially damped  by the baryonic chemical potential together with the gap  formed by the quarks that are charged with respect to the rotated electromagnetisms inside the superconductor.

\subsection{$C_V$ of Goldstone modes in the MCFL phase}

In the MCFL phase, the symmetry-breaking pattern is different from that in the CFL phase. 
Once a magnetic field is switched on, the difference between the
electric charge of the $u$ quark and that of the $d$ and $s$
quarks reduces the original flavor symmetry of the theory and
consequently also the symmetry group remaining after the diquark
condensate is formed. Then, the breaking pattern for the
MCFL-phase \cite{MCFL} becomes
\begin{equation}\label{MCFL}
\mathcal{G_{B}}=SU(3)_C \times SU(2)_L \times SU(2)_R \times
U(1)^{(1)}_A\times U(1)_B  \rightarrow
SU(2)_{C+L+R} 
\end{equation}
 In this case only five
Goldstone bosons remain. Three of them correspond to the breaking
of $SU(2)_A$, one to the breaking of $U(1)^{(1)}_A$, and one to
the breaking of $U(1)_B$. Thus, an applied magnetic field reduces
the number of Goldstone bosons in the superconducting phase, from
nine to five.

The MCFL phase is not just characterized by a smaller number of
Goldstone fields, but by the fact that all these bosons are
neutral with respect to the rotated electric charge \cite{Magnetic-Phases}. Thus, no
charged low-energy excitation can be produced in the MCFL phase.

Hence, we could think that once a magnetic field is
present, the original symmetry group $\mathcal{G}$ is reduced, to
$\mathcal{G_{B}}$, and that the low-energy theory
should immediately correspond to the breaking pattern [Eq. (\ref{MCFL})], having only five neutral Goldstone bosons. However, it is clear
that in very weak magnetic fields the symmetry of the CFL phase
can be treated as a good approximated symmetry, meaning that in
weak fields the low-energy excitations are essentially governed by
ten approximately massless scalars (those of the breaking pattern
(\ref{CFL})) instead of five. At this point, it is necessary to exactly understand what is the threshold field
strength that effectively separates the CFL low-energy behavior from that of
the MCFL.

The  threshold field that marks the reduction in the number of Goldstones was found in Ref.  \cite{Magnetic-Phases} and it is given by the following facts: As is known, a magnetic field when interacting with a charged boson endows it with an effective mass. Now, for a meson to be stable in this medium, its mass
should be less than twice the gap; otherwise, it will decay into a particle-antiparticle pair. This means that there exists a threshold field for the $CFL \rightarrow MCFL$ symmetry transition to be effective. In Ref. \cite{Magnetic-Phases} it was found to be of order $\sim 10^{16}$ G.

In the MCFL phase, where only neutral Goldstone bosons exist, four massive (with masses that do not depend on the magnetic field) and one massless (the one associated with the baryonic symmetry breaking), we have that the corresponding heat capacities are expressed by similar formulas to Eqs. (\ref{Massive-G}) and (\ref{Massless-G}), respectively. Thus, we conclude that in the MCFL phase also the contribution of the Goldstone mode associated with the breaking of the baryonic symmetry is the dominant one for the heat capacity.

\section{$C_V$ of neutron stars at intermediate density}

It has been shown in several QCD-inspired NJL models, that with increasing chemical potential, the energy separation between quarks and antiquarks increases up to a level where it is not energetically favorable anymore to excite antiquarks all the way from the Dirac sea to be paired with the quarks at the Fermi surface. In this case, various possibilities come into sight. Either no condensate is favored, and the chiral symmetry is restored; quarks and holes near the Fermi surface pair with parallel momenta, giving rise to inhomogeneous chiral condensates; or quarks may pair with quarks through an attractive channel at the Fermi surface to form a color superconductor phase that, at those moderate densities will preferably be also inhomogeneous \cite{Inh-CS}. The last two possibilities are usually favored \cite{Klimenko, Inh-CS, Israel}, meaning that the transition to a chirally restored phase found in NJL models with increasing density \cite{Klevansky}, will most likely occur in more than one step, or perhaps will not occur.

In this section, we want to investigate the heat capacity of the inhomogeneous chiral condensate phase, called the MDCDW phase, which is a chiral density wave phase with a quark-hole inhomogeneous condensate that forms at intermediate densities in the presence of a magnetic field \cite{Klimenko}. Such a phase is particularly interesting because of its anomalous electromagnetic transport properties \cite{MDCDW}  and its compatibility with various astrophysical constraints \cite{Estefano}.

We should point out that the inclusion of the magnetic field in this phase, apart from the fact that it is a natural ingredient in the astrophysics of NS, is necessary to secure the thermal stability of this phase. As it was demonstrated in \cite{MDCDW-2}, the presence of the magnetic field eliminates the so-called  Landau-Peierls instability \cite{Landau-Peirls Inst}, which is common to all  single-modulated condensate systems, meaning that in three spatial dimensions and in the absence of a magnetic field, the condensate is unstable against thermal fluctuations at any temperature value. But the explicit breaking of the rotational and isospin symmetries by the external magnetic field, together with the induction of a nontrivial topology that manifests itself in the asymmetry of the LLL modes, preserve the long-range order of the MDCDW phase at finite temperature.

\subsection{$C_V$ of quark matter in the MDCDW phase}

The temperature-dependent thermodynamic potential for this phase is given by \cite{Klimenko}
\begin{equation}
     \Omega^{MDCDW}_\beta=-\sum_{f=u,d}\frac{|e_fB|N_c}{(2\pi)^2\beta}\int_{-\infty}^{\infty}dp \sum_{l\xi\epsilon}\ln \left(1+e^{ -\beta(|E^f_{l,\xi,\epsilon}-\mu|)}\right)
     \label{Th-Pot-MDCDW}
     \end{equation}
where $N_c$ is the color number, $f$ denotes the flavor index for quarks $u$ and $d$, $l$ the Landau level number, and the energy spectra are given by
\begin{eqnarray} \label{MDCDW-E}
   E_{0,\epsilon}&=&\epsilon \sqrt{m^2+p^2}+b, \quad \epsilon=\pm, \l=0 \nonumber
\\
E^f_{l,\xi,\epsilon}&=&\epsilon \left [ \left (\xi\sqrt{m^2+p^2}+b \right )^2+2|e_f B| l \right ]^{1/2}, \quad \epsilon=\pm, \xi=\pm, l=1,2,3,...
\end{eqnarray}

In Eq. (\ref{MDCDW-E}), $m$ and $b$ are the amplitude and modulation of the inhomogeneous condensate, respectively. These are dynamical parameters that have to be found through the system gap equations \cite{Klimenko, Israel}. Note that the energy spectrum of the LLL is asymmetric. This asymmetry has topological implications as discussed in Refs. \cite{MDCDW, Tatsumi}.

From Eq. (\ref{Th-Pot-MDCDW}), we obtain the corresponding heat capacity
\begin{equation}
  C^{MDCDW}_V =\sum_{f=u,d} \frac{|e_fB|N_c}{(2\pi)^2}\sum_{l,\xi,\epsilon}\int_{-\infty}^{\infty}dp  \left( \frac{ |E^f_{l,\xi,\epsilon}-\mu |}{2T} \right ) ^2 sech^2 \left( \frac{ |E^f_{l,\xi,\epsilon}-\mu |}{2T} \right ) 
  \label{CV-MDCDW}    
\end{equation}

In the low-temperature limit we have
\begin{equation}
  C^{MDCDW}_V \simeq \sum_{f=u,d} \frac{4|e_fB|N_c}{(2\pi)^2}\sum_{l,\xi,\epsilon}\int_{-\infty}^{\infty}dp  \left( \frac{ |E^f_{l,\xi,\epsilon}-\mu |}{2T} \right )^2 e^{ - |E^f_{l,\xi,\epsilon}-\mu |/T }   
\end{equation}

Because of the negative exponential, we have that the leading contribution comes from the LLL, and  from $\epsilon =-$ because of the modulus in the exponent. Then, we can write Eq. (\ref{CV-MDCDW}) as 
\begin{eqnarray}
  C^{MDCDW}_V &\simeq &\sum_{f=u,d} \frac{2|e_fB|N_c}{(2\pi)^2}\int_{0}^{\infty}dp  \left( \frac{ |E^f_{0,-}-\mu |}{2T} \right ) ^2 sech^2 \left( \frac{ |E_{0,-}-\mu |}{2T} \right )
\\
&=&\sum_{f=u,d} \frac{2|e_fB|N_c}{(2\pi)^2}\int_{0}^{\infty}dp  \left( \frac{ |-\sqrt{p^2+m^2}+b-\mu|}{2T} \right ) ^2 sech^2 \left( \frac{ |-\sqrt{p^2+m^2}+b-\mu|}{2T} \right )  \nonumber
  \label{CV-MDCDW-LLL}    
\end{eqnarray}

Now, making the variable change $p/T \to p'$, we have
\begin{equation}
 C^{MDCDW}_V \simeq \sum_{f=u,d} \frac{2|e_fB|N_cT}{(2\pi)^2}\int_{0}^{\infty}dp'  \left( \frac{ |-\sqrt{p'^2+\hat{m}^2}+\widehat{b}-\widehat{\mu}|}{2} \right ) ^2 sech^2 \left( \frac{ |-\sqrt{p'^2+\widehat{m}^2}+\widehat{b}-\widehat{\mu}|}{2} \right )
 \label{Variable-Change-1}    
\end{equation}
The normalization notation $\hat{Q}=Q/T$ is used in the previous expression.

Introducing now the new variable change, 
\begin{equation}
 x=-\sqrt{p'^2+\hat{m}^2}+\widehat{b}-\widehat{\mu}=\hat{E}_{0,-}-\widehat{\mu}
  \label{Variable-Change-2}    
\end{equation}
with
\begin{equation}
 dp'=\frac{x-\widehat{b}+\widehat{\mu}}{\sqrt{(x-\widehat{b}+\widehat{\mu})^2-\widehat{m}^2}}  dx 
 \label{Variable-Change-3}    
\end{equation}
we obtain
\begin{equation}
 C^{MDCDW}_V \simeq \sum_{f=u,d} 2N_cT\int_{\widehat{m}+\widehat{b}-\widehat{\mu}}^{\infty}dx g_f(x) \left( \frac{ |x|}{2} \right ) ^2 sech^2 \left( \frac{ |x|}{2} \right )
 \label{Variable-Change-2}    
\end{equation}
where $g_f(x)$ represents the system density of states per unit volume given by
\begin{equation}
  g_f(x)=\frac{|e_fB|}{4\pi^2}
\frac{ x-\widehat{b}+\widehat{\mu}}{\sqrt{(x-\widehat{b}+\widehat{\mu})^2-\widehat{m}^2}}
\label{Func-Stat}    
\end{equation}

As was discussed in Section II-A, the Fermi-Dirac distribution at low temperature
only changes significantly in the vicinity of the Fermi energy. Hence, the integral in (\ref{Variable-Change-2}) mainly gets significant
contributions in the surroundings of $E_{0,-}=\mu$. Then, from (\ref{Variable-Change-2}) we see that $g_f(E_{0,-}=\mu) \equiv g_f(x=0)$. Thus,
we can approximate the density of states $\hat{g}_f(x)$ by its value at $\hat{g}_f(x=0)$. After this change and taking into account that in the $T \to 0$ limit the lower limit in the integral of Eq. (\ref{Func-Stat}) goes to negative infinity because $\mu > m, b$, we have

\begin{eqnarray}
  C^{MDCDW}_V &\simeq& 2N_cT\sum_{f=u,d}g_f(0) \int_{-\infty}^{{\infty}}\left(\frac{x}{2}\right)^2 sech^2 \left(\frac{x}{2} \right ) dx \nonumber
\\
&\simeq& \frac{2\pi^2N_c}{3}T\sum_{f=u,d}g_f(0) \nonumber
\\
&\simeq& \frac{(\mu-b)|eB|T}{2\sqrt{(\mu-b)^2-m^2}}
  \label{CV-MDCDW-LLL-3}    
\end{eqnarray}

In this case, since we neither have a superfluid, nor a superconducting state, the heat capacity is not as damped as in the CFL quark-matter system. The $C_V$ in this case depends on the dynamical parameters $m$ and $b$ that have to be determined by the minimum equations of the theory \cite{Klimenko}. These parameters will decrease with the temperature, but even at zero temperature, we have that $\mu > b > m$ in the region of interest \cite{Klimenko, Israel}. Hence, $C^{MDCDW}_V\sim eBT/2$. 

If we compare Eq. (\ref{Cv-LLL}) after neglecting $m_e$, with  Eq. (\ref{CV-MDCDW-LLL-3}) after neglecting $m$ and $b$, we see that the quark contribution to the heat capacity is 3 times larger than the electron contribution at the same temperature and magnetic field. This is a consequence of the fact that in this quark phase we have more degrees of freedom, especially  the three color degrees of freedom, because the flavor ones enter in $C_V$ through $|e_fB|$; henceforth, the fractional quark charges in the flavor sum  give $|eB|$ without an extra factor.

We should mention that at intermediate densities, there are other inhomogeneous phases that can compete with the MDCDW phase. With decreasing density, the combined effect of the strange quark mass, neutrality constraint and beta equilibrium tends to pull apart the Fermi momenta of different flavors in the CFL phase, imposing an extra energy cost on the formation of Cooper pairs. For the two-flavor 2SC color superconductor, the neutrality and beta equilibrium conditions also produce a mismatch in the Fermi spheres of different flavors. BCS pairing then dominates as long as the energy cost of forcing all species to have the same Fermi momentum is compensated by the pairing energy that is released by the formation of Cooper pairs. Therefore, with decreasing density, homogeneous CS phases like the CFL and the 2SC become gapless and, most importantly, become unstable \cite {Gatto, Huang}.The instability, known as chromomagnetic instability, manifests itself in the form of imaginary Meissner masses for some of the gluons and indicates an instability towards spontaneous breaking of translational invariance \cite{Rupak}. In other words, it indicates the formation of a spatially inhomogeneous phase. 

Most inhomogeneous CS phases are based on the ideas of Larkin and Ovchinnikov (LO) \cite{Larkin} and Fulde and Ferrell (FF) \cite{Fulde}, originally applied to condensed matter. In the CS LOFF phases \cite{Bowers, Nardulli}, quarks of different flavors pair even though they have different Fermi momenta, because they form Cooper pairs with nonzero momentum. CS inhomogeneous phases with gluon condensates that break rotational symmetry \cite{Incera-2} have also been considered to solve the chromomagnetic instability. 
These inhomogeneous phases are expected to have large $C_V$ values because of the presence of gapless fermionic modes. But its study is still a pending task.

\section{Heat Capacity estimates for  different neutron star compositions}

In this section, we will consider the main contributions to the heat capacity of the different phases and compare them with the phenomenological constraint $\tilde{C}_V\gtrsim 10^{36}(T/10^8)$ erg/K.

\subsection{Nuclear matter phase}

In the relatively low-density phase, the main contribution to the heat capacity comes from neutrons, protons and electrons. When neutrons and protons are forming superfluid and superconducting states respectively, their contributions to $C_V$ become negligible, as we already discussed. Thus, the contributions of unpaired neutrons, together with those of electrons, become the leading ones. As follows, by making a rough approximation, we will estimate the order of the contributions of the main participants to the heat capacity of NS in the absence and in the presence of a magnetic field.

\subsubsection{Unpaired neutrons}

The calculation of the heat capacity of neutrons is very similar to that of electrons, only with the change of the electric chemical potential $\mu_e$ by the baryonic chemical potential $\mu$ and of the electron mass $m_e$ by the neutron mass $m_n$. Thus, making these replacements in  Eq. (\ref{Cv-electron}), we obtain
 \begin{equation}
    C_V^n\simeq \frac{ \mu T}{3}\sqrt{\mu^2-m_n^2}
    \label{Cv-neutron}
\end{equation}

Taking into account that the Fermi momentum is  given by $k_F=c\hbar[(3\pi^2/2)n_N]^{1/3}$; for $n_B=3n_s$, where $n_s=015 fm^{-3}$ is the saturation density, the corresponding baryonic chemical potential is $\mu=1,009$  MeV and the Fermi temperature $T_F=\mu/k_B=11.7 \times 10^{12}$ K.

Considering that $\mu > m_n$, Eq. (\ref{Cv-neutron}) can be simplified as
\begin{equation}
    C^{N}_V\simeq \frac{ \mu^2k^2_BT}{3},
    \label{Cv-neutron-2}
\end{equation}
where we include the square of the Boltzmann constant, associated with the second derivative with respect to $T$ in the definition of the heat capacity [Eq. (\ref{sh})] (Notice, on the other hand, that in the thermodynamic potential, $T$, appears always multiplied by $k_B$.) Taking into account that the neutron number density is given in this approximation by $n_N=2\mu^3/3\pi^2$ and that the Fermi temperature is given by $k_BT_F=\mu$, we can rewrite Eq. (\ref{Cv-neutron-2})
as
\begin{equation}
    C^{N}_V\simeq\frac{ \pi^2}{2} n_Nk_B\left(\frac{ T}{T_F}\right)
    \label{Cv-neutron-3}
\end{equation}

Then, for $n_N\simeq 3n_s$, and $T=10^8$ K, we obtain the volumetric heat capacity
\begin{equation}
    C^{N}_V\simeq 0.3 \times 10^{19} erg/K.cm^3
    \label{Cv-neutron-4}
\end{equation}

To compare with the results given in  \cite{Cv-NS} for the heat capacity, we will multiply $C^{N}_V$ by the volume of a NS of radius $10$ Km ($V_{NS}=(4/3)\pi R^3=4.2 \times 10^{18} cm^3$). Although it is a very rough approximation that does not take into account all the other components known to exist in the NA \cite{Page}, or the fact that the particle density is not uniform in the star interior, it can be used to find the order of the heat capacity for a given temperature
\begin{equation}
   \tilde{C}^{N}_V =  C^{N}_V\times V_{NS}=0.1\times10^{38} erg/K,
    \label{Cv-neutron-5}
\end{equation}
which is of the same order as the one reported in Ref. \cite{Cv-NS}.

The inclusion of a magnetic field can be done through the anomalous-magnetic-moment/magnetic-field interaction. The calculation of $C_V$ for neutrons in a magnetic field was performed in Ref. \cite{Aric} and it was found that 
up to magnetic fields of $10^{18}$ G, the magnetic-field effect on $C_V$ is negligibly small.

The calculation for unpaired protons at $B=0$ is very similar to the one done for neutrons with the replacement $n_P=X_en_N$,  where $n_P$ and $n_N$ are the number densities of protons and neutrons, respectively, and  $X_e$ is the electron fraction, which is expected to be larger than a threshold value $X_e\sim 1/9$ in order for the direct URCA process to occur in NS \cite{Prakash}. In the presence of a magnetic field, the calculation will be similar to that fused for electrons below in Section V-A3, with the replacement of the corresponding masses, $m_e\rightarrow m_n$,  and chemical potentials $\mu_e\rightarrow \mu-\mu_e$.

\subsubsection{Electrons at $B=0$}

We start from Eq. (\ref{Cv-electron}), written in the form
\begin{equation}
    C^{e}_V\simeq \frac{ \mu_e^2k^2_BT}{3},
    \label{Cv-electron-2}
\end{equation}
where we neglect $m_e$ under the consideration that $\mu_e>m_e$, as shown in the Appendix, and we  include the square of the Boltzmann constant as in Eq. (\ref{Cv-neutron-2}).

Taking into account that the electron number density is given in this approximation by $n_e=2\mu_e^3/3\pi^2$ and that the Fermi temperature is given by $k_BT_F=\mu_e$, we can write Eq. (\ref{Cv-electron-2})
as
\begin{equation}
    C^{e}_V\simeq \frac{\pi^2}{2} n_ek_B\left(\frac{ T}{T_F}\right)
    \label{Cv-electron-3}
\end{equation}

To estimate the electron number density, we take into account that  due to the electric neutrality $n_e=n_P=X_e n_N$. Thus, for $n_N=3n_s$, we have $n_e=(1/3)n_s$, with a corresponding $T_F=2.1 \times 10^{12}$ K. Substituting with these  values in Eq.  (\ref{Cv-electron-3}), we obtain for $T=10^8$ K the volumetric heat capacity
\begin{equation}
    C^{e}_V\simeq 0.2 \times 10^{19} erg/k.cm^3
    \label{Cv-electron-4}
\end{equation}

Multiplying by the volume for a NS with a $10$ km radius, as we did for neutrons, we obtain
\begin{equation}
\tilde{C}^{e}_V =  C^{e}_V\times V_{NS}=0.8\times10^{37} erg/K,
    \label{Cv-electron-5}
\end{equation}
which is of the same order as the value reported in Ref. \cite{Cv-NS}.

We should call attention to the fact that for massless Goldstones, $\tilde{C}^{G}_V \sim n_s(T/T_F)^3 \ll \tilde{C}^{e}_V $ since $(T/T_F)\sim 10^{-4}$.

\subsubsection{Electrons at $B\neq0$}

Using the same approximation as in the previous section, the electron heat capacity in the presence of a magnetic field (\ref{Cv-LLL}) can be written as
\begin{equation}
     C_{V}^{e}(B) \simeq \frac{eBk^2_BT}{6}
\label{Cv-LLL-2}
\end{equation}

The electron number at zero temperature in this case is given by
\begin{equation}
    N^{e}(B)=-\frac{\partial \Omega^{(e)}_{LLL}}{\partial \mu_e}=\frac{eB}{2\pi^2}\sqrt{\mu_e^2-m_e^2}\simeq\frac{eB\mu_e}{2\pi^2}
\label{Ne-LLL}
\end{equation}
Hence, we see that to have the same number density of electrons as in the $B=0$ case, it is necessary, for the same chemical potential to experience a magnetic field of $B=2\mu^2/3=1.7\times10^{19}$ G, which is a value too high to be expected in the NS core \cite{Aric}.

Multiplying and dividing the rhs of Eq. (\ref{Cv-LLL-2}) by $\mu_e$ and introducing the Fermi temperature as before we have
\begin{equation}
     C_{V}^{e}(B) \simeq \frac{\pi^2N_{(LLL)}^{(e)}k_B}{3}\left(\frac{T}{T_F}\right)
\label{Cv-LLL-3}
\end{equation}

Keeping the same parameters as in the $B=0$ case, we obtain
\begin{equation}
    \tilde{C}_{V}^{e}(B) \simeq 0.2
    \times10^{37} erg/K,
\label{Cv-LLL-3}
\end{equation}
which is of the same order  than $C^{(e)}_V $ for magnetic fields of order $10^{19}$ G, but if we consider a more realistic field value of $B \sim 10^{17}$ G, the heat capacity decreases in 2 orders. Hence, we find that in the presence of  a moderately high magnetic field $\sim 10^{17}$ G, the electron heat capacity decreases.

\subsection{Quark phase}

Here we consider the contribution of quark matter at intermediate densities and in the presence of a magnetic field to the heat capacity of NS. At those densities, an interesting and promising candidate is the MDCDW phase.

In our analysis, we consider a quark star, which is formed only by quarks and a small cloud of electrons occupying a region of a few fm around the quark surface \cite{Alcock}. It is believed that the most energetically favored configuration for a NS, once quark matter is present, is one where the star is purely formed by three-flavor ($u$, $d$ and $s$) or two-flavor ($u$ and $d$) quark matter, giving rise to what is called a quark star. The possibility of having the three-flavor configuration is based on the Bodmer-Terazawa-Witten hypothesis \cite{BTW}, which proposes that  the whole NS will likely be converted into strange matter. The reason is that strange matter, which consists of roughly equal numbers of $u$, $d$, and $s$ quarks at high densities, is absolutely stable (it has lower energy per baryon than ordinary iron nuclei). More recently, in Ref. \cite{Holdom}, by using a phenomenological quark-meson model that includes the flavor-dependent feedback of the quark gas on the QCD vacuum, it was demonstrated that $u$-$d$ quark mater is in general more stable than strange quark matter, and it can be more stable than the ordinary nuclear matter when the baryon number is sufficiently large above  $A_{min}\gtrsim 300$, which, on the other hand, ensures the stability of ordinary nuclei. The two-flavor quark star is the one we are considering to realize the MDCDW phase in this section. In this case, the NS interior will be formed only by quarks.

\subsubsection{ Quarks in the MDCDW phase}

As discussed in Section IV, in the range of intermediate densities where the realization of the MDCDW phase is feasible, the heat capacity [Eq. (\ref{CV-MDCDW-LLL-3})]  can be reduced to 
\begin{equation}
     C^{MDCDW}_V\simeq 2eBT
\label{CV-MDCDW}
\end{equation} 

To follow a similar method to previous cases, we need to consider now the quark number density for this phase. This was calculated in Ref. \cite{MDCDW}, giving two components 
\begin{equation}
    n_{q}=n_{anom}+n_{ord}
\label{MDCDW-N}
\end{equation}
where $n_{anom}=3|eB|b/2\pi^2$ is the so-called anomalous quark number density, and $n_{ord}=3|eB|\mu/2\pi^2$ is the ordinary contribution coming from the MDCDW noanomalous many-particle thermodynamic potential at zero temperature.
Since $\mu > b$ in the region of interest, we have that the leading contribution comes from the noanomalous part, 
\begin{equation}
    n_{q}\simeq n_{ord}=\frac{3|eB|\mu}{2\pi^2}
\label{MDCDW-N-2}
\end{equation}
Then, substituting with Eq. (\ref{MDCDW-N-2}) in Eq. (\ref{CV-MDCDW}), we obtain
\begin{equation}
     C^{MDCDW}_V \simeq \frac{4\pi^2}{3}n_{q}k_B \left(\frac{T}{T_F}\right)
\label{V-3}
\end{equation}

To find the quark number density, let us consider that the baryonic number density is $n_B=3n_s$; thus, the quark number density is $n_q=n_B/3=n_s$, and the corresponding Fermi temperature is $T_F=2.9\times 10^{12}$ K. 
Substituting these quantities in Eq. (\ref{V-3}), we obtain for $T=10^{8}$ K the volumetric heat capacity
\begin{equation}
    C^{MDCDW}_V\simeq 0.9 \times 10^{19} erg/K.cm^3
    \label{Cv-neutron-6}
\end{equation}

Multiplying by the volume of a star of radius 10 km, we find the corresponding heat capacity given by
\begin{equation}
   \tilde{C}^{MDCDW}_V =  C^{N}_V\times V_{NS}=0.4\times10^{38} erg/K,
    \label{Cv-neutron-7}
\end{equation}
which is of the same order as the one obtained for unpaired neutrons [Eq. (\ref{Cv-neutron-5})].

\section{Concluding Remarks}

Compact star-cooling places some of the strictest constraints on determining the stellar inner phase \cite{Rafelt}.  In this sense, the heat capacity of the matter forming the core on certain external conditions of density, temperature, and applied magnetic field plays a fundamental role. In this paper, we have investigated the leading contribution to the heat capacity of different matter states that can exist in thelow-, intermediate-, and high-density star regions at low temperature and in the presence of a magnetic field, with the main goal to ascertain if quark matter can play a role in the star's inner composition. 

Long-term observations of NS temperatures in the range  from months to years after the accretion outburst has given information about the crust heat capacity \cite{Crust}, and continued observations on timescales of years can shed light on the limiting value of the heat capacity of the star core. Following this last search, in Ref. \cite {Cv-NS} it was found that there exists a lower limit to NS core heat capacity. This lower-limit value [$C_V\gtrsim 10^{36}(T/10^8)$ erg/K] put out of the game the matter components that exhibit superfluidity or superconductivity of any kind, since as we showed here, all of these cases are exponentially damped, so only electrons and unpaired neutrons will be able to satisfy the constraint. This will lead to the striking conclusion that, if the only quark-matter state to be realized in NS interior is the color superconducting CFL/MCFL phases, quarks will be ruled out from forming part of the NS core \cite {Cv-NS}. 

Nevertheless, as we already pointed out, the CFL/MCFL phases are the energetically favored phases at asymptotically high densities, which is barely expected to be reached in the inner core of NS. At intermediate densities, other phases can be realized \cite{Buballa}, and specifically chiral inhomogeneous condensate phases formed by particle-hole pairs are suitable candidates. In this regard we have considered in this paper the MDCDW phase and proved that at low temperatures its heat capacity is a linear function of temperature, as is also the case for electrons and unpaired neutrons. 

Another ingredient we considered in this investigation is the effect of the magnetic field on $C_V$. As known, significantly high magnetic fields populate NS. The magnetic field in the two cases we studied, which did not present exponential damping  (i.e. electrons and quarks in the MDCDW phase), makes two main contributions: one, redefining the density of states that depends on $eB$, and two, producing the Landau quantization of the energy levels. When considering the low-temperature limit, which is the proper one to describe the thermodynamic of NS once $k_BT\ll E_F$, we found that the main contribution to $C_V$ comes from the LLL in both systems. The explanation for that becomes clear by taking into account that at those low temperatures the particles lying on the Fermi sphere do not have enough energy to jump over the gap separating the Landau energy levels, which are of order $\sqrt{eB}> T$. In the case of the MDCDW phase, the magnetic field in addition has particular importance, since it is responsible of the stability of the phase under thermal fluctuations \cite{MDCDW-2}.

The chemical potential affecting the electron Dirac distribution, is only the electric chemical potential, while the protons have in addition the baryonic chemical potential. Taking into account the neutrality condition for the stellar medium formed by protons and electrons as charged particles, we find in Fig. 1 of the Appendix that $\mu>\mu_e>m_e$. Hence, neglecting $m_e$ in  Eqs. (\ref{Cv-electron}) and (\ref{Cv-LLL}) is a good approximation. It is also found that the presence of a magnetic field decreases the electron contribution to the NS heat capacity. To have the same density of electrons in the absence and in the presence of a magnetic field it is necessary for a baryonic density of $3n_s$ to experience a magnetic field of $B=2\mu^2/3=1.0\times10^{19}$ G, which is an extremely high value for the NS core \cite{Aric, Cardall}. For baryonic densities of that order and magnetic fields of smaller values, the NS heat capacity in the presence of a magnetic field decreases as compared with the corresponding one at zero magnetic field.

Finally, comparing the contributions to NS heat capacity of electrons in a magnetic field and quarks in the MDCDW phase, we find that the quarks have a contribution 1 order larger than electrons and 2 orders larger than the low-limit constraint found in Ref. \cite{Cv-NS} (see Section V).  Moreover, to guarantee the electric neutrality of the MDCDW phase, since the two quark flavors cannot cancel out in any combination the total electric charge of the system, we also need electrons that will contribute as well to the NS heat capacity \cite{Estefano}. Therefore, we conclude that quarks cannot be ruled out from the inner composition of NS, since if the MDCDW phase can be realized there, their corresponding contribution to the NS heat capacity will be even higher than that of electrons.

\acknowledgments
This work was supported in part by NSF grant No. PHY-2013222. 

\appendix

\section{Neutrality condition for an electron-proton system}

It is an accepted fact that NS are electrically neutral. Because of the strength hierarchy between the electromagnetic and the gravitational forces, the net charge should be extremely small, with $Z_{net}\sim |Z_p - Z_e|< 10^{-36} A$, where $Z_p$ and $Z_e$ are the proton and electron numbers and $A$ the baryon number \cite{Glendenning}. 

Our goal is to find the relation between the baryonic chemical potential, $\mu$, the electric chemical potential, $\mu_e$ and the electron mass, $m_e$, by using the condition that the NS medium iremains electrically neutral.

We consider here that the density is not high enough to give rise to other charged degrees of freedom different from protons and electrons, as it could be more massive leptons or hyperons. 
Thus, the global charge neutrality condition is given by
\begin{equation}
    \textit{Q}_{net} =Q_e+Q_p=-e\textit{N}_{e}+eN_p=e\left ( {\partial_{\mu}}_e \Omega_e -{\partial_{\mu}}_p \Omega_p \right )=0
    \label{neq}
\end{equation}
where the proton chemical potential is given by $\mu_p=\mu-\mu_e$.
Here the thermodynamic potentials for electrons and protons are given, respectively, by
\begin{equation}
     \Omega_e = -\frac{2}{\beta}\int_{-\infty}^{\infty}\frac{d^3p}{(2\pi)^3}\ln\left[(1+ e^{-\beta(\epsilon_e+\mu_e)})(1+e^{-\beta(\epsilon_e-\mu_e)})\right]
\end{equation}
\begin{equation}
     \Omega_p = -\frac{2}{\beta}\int_{-\infty}^{\infty}\frac{d^3p}{(2\pi)^3}\ln\left[(1+e^{-\beta(\epsilon_p+\mu_p)})(1+e^{-\beta(\epsilon_p-\mu_p)})\right]
\end{equation}

\begin{figure}[h!]
    \centering
    \includegraphics[scale=0.8]{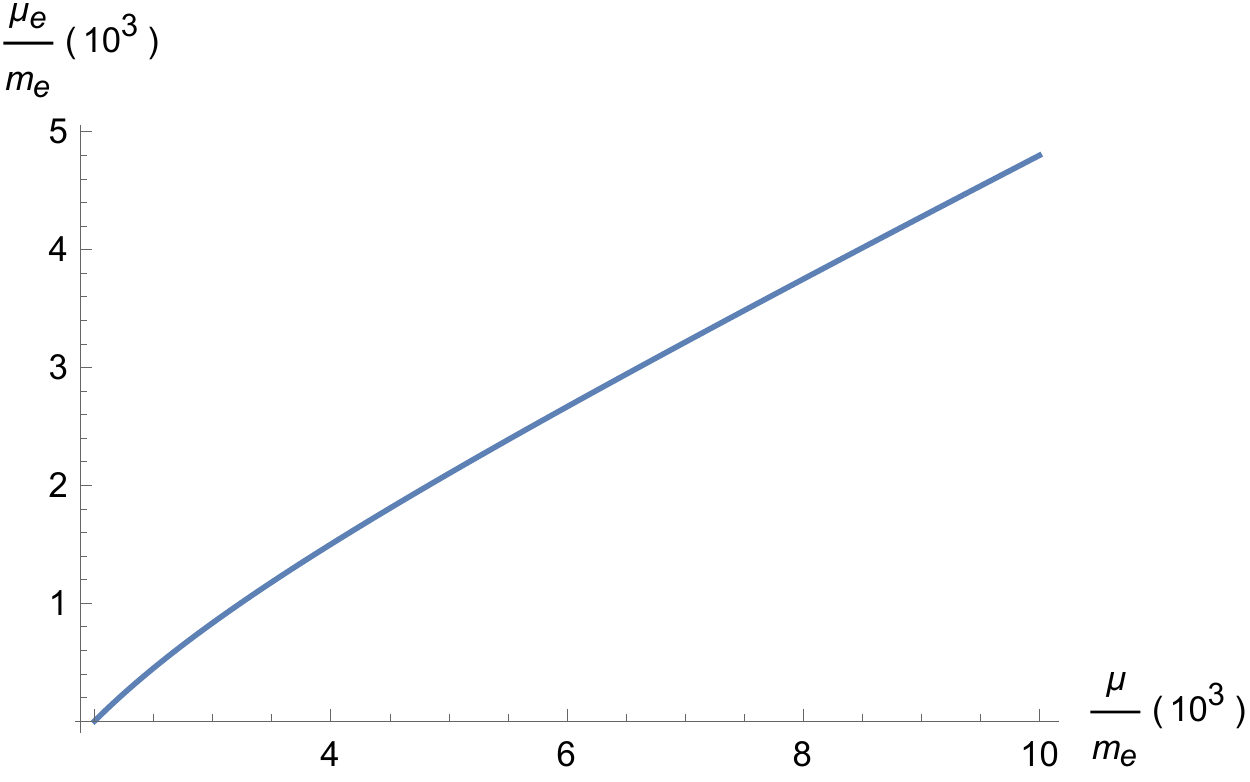}
    \caption{Dependence of $\mu_e/m_e$ on $\mu/m_e$. We can observe that $\mu > \mu_e > m_e$.}
    \label{neqplot}
\end{figure}

The corresponding particle numbers are given, respectively, by
\begin{equation}
    N_e=-{\partial_\mu}_e \Omega_e = \frac{1}{\pi^2}\int_0^\infty p^2dp\left(\frac{1}{(1+e^{\beta(\epsilon_e-\mu_e)})}-\frac{1}{(1+e^{\beta(\epsilon_e+\mu_e)})}\right)
\end{equation}
\begin{equation}
    N_p=-{\partial_\mu}_p \Omega_p = \frac{1}{\pi^2}\int_0^\infty p^2dp\left(\frac{1}{(1+e^{\beta(\epsilon_p-\mu_p)})}-\frac{1}{(1+e^{\beta(\epsilon_p+\mu_p)})}\right)
\end{equation}

In the zero-temperature limit and considering that $\mu_e >0$ and $\mu_p >0$, and taking into account the electron and proton dispersion relations,  $\epsilon=\sqrt{p^2+m_e^2}$ and $\epsilon_p=\sqrt{p^2+m_p^2}$, respectively,  we have
\begin{equation}
   N_e =\frac{1}{\pi^2}\int_0^\infty p^2dp \theta\left(\epsilon_e-\mu_e\right)= \frac{1}{\pi^2}\int_0^{\sqrt{\mu_e^2-m_e^2}} p^2dp = \frac{1}{3\pi^2}\left [\mu_e^2-m_e^2 \right ]^{3/2}
   \end{equation}
\begin{equation}
   N_p=\frac{1}{\pi^2}\int_0^\infty p^2dp \theta\left(\epsilon_p-\mu_p\right)= \frac{1}{\pi^2}\int_0^{\sqrt{\mu_p^2-m_p^2}} p^2dp = \frac{1}{3\pi^2}\left [\mu_p^2-m_p^2 \right ]^{3/2}
\end{equation}

Imposing the neutrality condition, $N_e=N_p$, that follows from (\ref{neq}), we obtain
\begin{equation}
    (\mu_e-\mu)^2-m_p^2 =\mu_e^2-m_e^2
    \label{neutrality}
\end{equation}

For $m_e=0.5$ MeV and $m_p=938.3$ MeV, and introducing the notation, $x=\mu/m_e$ and $y=\mu_e/m_e$, the neutrality condition (\ref{neutrality}), can be written as
\begin{equation}
    y = \frac{1-0.4\cdot10^7}{2x}+\frac{x}{2}
\end{equation}
 
 This function is plotted in Fig.1 for the domain of interest for NS.
 Since $y>0$, we have that $x>2\times10^3$, and we consider $\mu$ values corresponding up to ten times nuclear density.

\end{document}